\documentclass[twocolumn,epjc3]{svjour3}

\usepackage{tocloft,hyperref}

\usepackage{graphicx,url,booktabs}				% Use pdf, png, jpg, or eps§ with pdflatex; use eps in DVI mode
\usepackage{amsmath,amssymb}
\usepackage{comment,color}
\definecolor{verde}{rgb}{0.13, 0.55, 0.13}

\newcommand{\corr}[1]{{\leavevmode\color{black}#1}}
\newcommand{\corrca}[1]{{\leavevmode\color{black}#1}} %carlo's corrections
\newcommand{\agr}[1]{{\leavevmode\color{black}#1}}
\newcommand{\rvw}[1]{{\leavevmode\color{black}#1}}

\newcommand{\ele}{\mathrm{e}}
\newcommand{\prot}{\mathrm{p}}
\newcommand{\hh}{\mathrm{h}}
\newcommand{\wbos}{\mathrm{W}}
\newcommand{\neut}{\mathrm{n}}

\makeatletter
\newcommand{\oset}[3][0ex]{%
\mathrel{\mathop{#3}\limits^{
\vbox to#1{\kern-2\ex@
\hbox{$\scriptstyle#2$}\vss}}}}
\makeatother
\newcommand\mysymb{\scalebox{.3}{(}\raisebox{-1.7pt}{$-$}\scalebox{.3}{)}}

\title{Introduction to neutrino astronomy}

\author{A.~Gallo Rosso\thanksref{a1,a2}
\and
C.~Mascaretti\thanksref{a1}
\and
A.~Palladino\thanksref{a3} 
\and
F.~Vissani\thanksref{a2}}

\institute{Gran Sasso Science Institute, Viale F.~Crispi 7, L'Aquila, Italy\label{a1} 
\and 
INFN--Laboratori Nazionali del Gran Sasso Science, Assergi (AQ), Italy\label{a2}
\and 
DESY, Platanenallee 6, 15738
Zeuthen, Germany\label{a3}}

\date{}							% Activate to display a given date or no date

\begin{document}

\maketitle

\begin{abstract}
{\small\sf This paper is an introduction to neutrino astronomy, addressed to astronomers and written by astroparticle physicists.
The focus is on achievements and goals in neutrino astronomy, rather than on the aspects connected to particle physics,
however the particle physics concepts needed to understand the peculiar neutrino features are also introduced.
The material is selected - i.e., not all achievements are reviewed - making however efforts to highlight current research issues.

}
\end{abstract}

%\tableofcontents

%\bigskip\bigskip\bigskip
%\section{}
%\subsection{}

%\section*{Introduction}

\bigskip

\noindent
%\paragraph{Outline:}
The layout of this paper 
is as follows: 
In Sect.~\ref{sec:nt}, we introduce the neutrinos, examine their interactions, and present \rvw{neutrino detectors and telescopes}.
In Sect.~\ref{sec:so}, we discuss solar neutrinos, that have been detected and are matter of intense (theoretical and experimental) studies. 
In Sect.~\ref{sec:sn}, we focus on supernova neutrinos, that inform us on a very dramatic astrophysical event and can tell us a lot on 
the phenomenon of gravitational collapse. 
In Sect.~\ref{sec:cn}, we discuss the highest energy \corrca{neutrinos,} a very recent and lively research field. 
In Sect.~\ref{sec:no}, we review the phenomenon of neutrino oscillations and assess its relevance for neutrino astronomy.
Finally, we offer a brief overall assessment and a summary in Sect.~\ref{sec:di}.
\rvw{In order to help the beginner, we prefer to limit the list of references,  
opting whenever possible for review works such as \cite{bahcall3,r1,r2,astro-book}.}

%\clearpage

\section{Neutrino properties and neutrino telescopes\label{sec:nt}}

\subsection{General properties}
Neutrinos are neutral particles \corrca{which interact very rarely} with matter; to have an idea of how small they are, if an atom were scaled to the size of the Earth (namely, if one zoomed in by a factor of about $10^{17}$) the atomic nucleus would have the size of a football field and solar neutrinos would have the size of one virus!
Evidently it is not easy to ``see'' them.

There are 3 types of neutrinos, traditionally called `flavors'.
They are identified by the charged particles that neutrinos can produce by interacting with matter. 
\corrca{Such} charged particles are the electron ($\ele$), the muon ($\mu$), the tau ($\tau$), all with the same negative charge $-1$, in units of $1.602\times 10^{-19}$ Coulombs.\footnote{
These particles differ greatly for the mass: the muon weighs $\sim 200$ times the electron, and the tau about 3,500 times.}
For instance, an electron neutrino $\nu_\ele$ is the neutral particle that, \corrca{when interacting}, can produce an electron; a muon neutrino $\nu_\mu$ can produce a muon, etc.
Likewise, antineutrinos can be defined as those particles that can produce the various particles with charge $+1$; e.g., a tau antineutrino $\bar\nu_\tau$ produces an anti-tau, with mass equal to the tau particle. 
Using the \corrca{formalism} of chemical reactions, we will write 
\begin{equation}
\nu_\ell + X \to \ell^- + Y \mbox{ with } \ell=\ele,\mu,\tau
\end{equation}
where $X$ is an initial and $Y$ a final particle (or set of particles), and their electric charges are $Q(Y)=Q(X) +1$. 
\corrca{Following the usual convention, we} indicate neutrinos by the symbol $\nu$, and antineutrinos by $\bar\nu$.
A brilliant pictorial summary of the main neutrino features can be found at the following web site: 
\begin{center}
\url{http://www.quarked.org/askmarks/answer4.html}
\end{center}

Note finally that, in the observations of neutrino astronomy, as a rule,  neutrinos should be considered as individual particles rather than waves. 
In comparison to ordinary astronomy, it is as if neutrino astronomy were always in the single-photon mode. 
Peculiar wave phenomena, however, affect the propagation of neutrinos (see  Sect.~\ref{sec:no}).

\subsection{Neutrino interactions}
\corrca{Physicists quantify the probability of any interaction (or process) to take place by computing the corresponding ``cross sections'', which can be interpreted as the transversal size of a target we would like to hit.}
The cross sections are usually indicated by the symbol $\sigma$. 
Quite in general, the cross section is linked to the 
\rvw{number density} 
of targets $n$ and to the mean free path $\lambda$ by the relation, $n\times \sigma \times \lambda=1$; if a particle 
travels  a distance $d,$ one calls opacity the product $\tau= n\times \sigma \times d$, that \corrca{quantifies} the probability of 
no interaction 
\begin{equation}P(d)=\exp(-\tau)=\exp(-d/\lambda)\end{equation}
The estimation of the size of the neutrino $R_\nu$ given in the previous section is obtained by setting $\sigma_\nu\equiv \pi R^2_\nu$, \corrca{where $\sigma_\nu$ is the typical cross-section of a process involving neutrinos.
Such processes have cross-sections which} are proportional to a universal constant
\begin{equation}
G_F^2=5.297\times 10^{-44}\mbox{ (cm/MeV)}^2
\end{equation}
where $G_F$ is called Fermi constant. 
\corrca{The vast majority of} neutrino cross sections are therefore given by $G_F^2$ times the square of an energy or the product of an energy and a mass, simply for dimensional reasons. 
However, the theory links the interactions of the neutrinos with other observable processes: for instance, all reactions and decay processes listed in table~\ref{tab:p} \corrca{share} the same dynamics. 
Note that all six reactions of table~\ref{tab:p} can be obtained one from the other by \corrca{either} exchanging a particle from one side to the corresponding antiparticle in the other side of the reaction, or exchanging the direction of the arrow. 
The decay rates $\Gamma$, in particular, are also proportional to $G_F^2$; again, for dimensional reasons, this is multiplied by the fifth power of a mass.\footnote{We are using the so called natural units. Setting $c=1$, masses and energies and momenta have the same dimension, and likewise time and space; setting $\hbar=1$, and due to the uncertainty principle $\Delta x \Delta p\ge \hbar/2$, space and momenta have inverse dimensions. Therefore, a rate (inverse time) has the dimensions of an energy.} 

\begin{table}[t]
%\begin{minipage}[c]{3.5cm}
	\caption{\small\em Most common reactions involving electronic neutrinos and antineutrinos with
	neutrons and protons ($n$ and $p$)  
	and their 
	traditional names: $\beta^\pm$ decay, electron (positron) capture,
	Inverse Beta Decay (IBD). }
%	\end{minipage}
	%\vskip2mm
    \centering
	\begin{tabular}{rr}
    name & reaction \\
    \midrule
	%\toprule
	$\beta^-$ decay 	& $\neut\to \prot +\ele^- +\bar\nu_\ele$	 \\
	$\beta^+$ decay 	& $\prot \to \neut + \ele^+ + \nu_\ele	$ \\
    $\beta^-$ capture 	& $\prot + \ele^-\to \neut  +\nu_\ele$	\\
	$\beta^+$ capture 	& $\neut + \ele^+\to \prot + \bar\nu_\ele$	\\ 
	IBD					& $\prot + \bar\nu_\ele \to \neut  + \ele^+$	\\
	IBD on $n$			& $\neut + \nu _\ele\to \prot + \ele^-$		\\
	%\botrule
\end{tabular}
\label{tab:p}
\end{table}

\subsection{Principles of neutrino \rvw{detectors and telescopes}}

In view of the small chance of seeing these particles,  it is quite evident that 
 neutrino detectors need to have a large amount of target particles.
 Let us indicate their number \corrca{by} $N_{\mbox{\tiny target}}$
 and consider an intense flux of neutrinos/antineutrinos, measured e.g., in $1/(\mbox{cm}^2\mbox{s})$, 
 that we call $\Phi_\nu$.
In the time of observation (or of emission)~$T$, we will have that the number of observable events is: 
\begin{equation}
\mathcal{N}_{\mbox{\tiny events}}=N_{\mbox{\tiny targets}}\times T\times \sigma\times \Phi_\nu
\end{equation}
where $\sigma$ is the relevant cross section. 
In view of the above considerations, and more precisely of the smallness of the cross section, only a few specific sources of neutrinos can be observed. 
Three of them will be examined in the following three Sections; for a more complete list, see \cite{astro-book,rev1}. 
Moreover, the number of events due to other processes, that can mimic neutrino-induced events (so called {\em background} processes) 
should be minimized.  
For this reason, \rvw{neutrino observatories} are located underground, underwater or under-ice, at a sufficient depth, so that the cosmic radiation \corrca{and its byproducts are} screened. 
On top of that, the neutrino detectors \corrca{which aim} at observing \corrca{low-energy (MeV)} neutrinos have to be sure that radioactive decays do not give an excessive amount of spurious events. 
In practice, this means that any neutrino detector has to restrict \corrca{its} observations only to those events that fall above a minimal energy (``threshold'') or that arrive at a certain time or from a certain direction of the sky. 

\rvw{Later, we will use the more specific term `neutrino telescope' rather than the generic one `neutrino detector' 
in the few cases when the detector can provide astronomers with information on the direction of arrival.
However, note that, every so often, the two terms are just synonyms, 
see e.g.,~\cite{syn}; conversely, from time to time, `neutrino telescope' is used in the narrow sense of 
`high energy neutrino telescope', see e.g.,~\cite{a5}.}

\section{Solar neutrinos\label{sec:so}}

\rvw{In this section, we introduce the theoretical description of the solar neutrinos 
and examine the current observational knowledge. 
A few references that can help to access more deeply into 
these topics are \cite{bahcall3,r1,r2,astro-book,r3}.}

%\subsection{Introduction}

%The Sun \corrca{emits} emits energy in the form of light but also neutrinos.
\rvw{The understanding of the reason of solar energy took a lot of time. 
One early idea (Perrin, 1919, \cite{pero}) was that, in   
the reaction 4H$\to$He among {\em atoms}, the decrease in mass of 
$\sim 1$\%  liberates energy. 
In actuality, this happens through various sequences 
of {\em nuclear} transformations, where the net disappearance of 2 
electrons is compensated by the appearance of 2 electron neutrinos.\footnote{\rvw{Symbolically,  we have 
$4\prot +2\ele^-\to \alpha+2\nu_\ele$.}}}

In the Sun, 
the most important of such 
reactions is the proton-proton fusion ($\rm D = $ $^2$H nucleus):
\begin{equation}
%\phantom{centramquestaeq} 
\prot + \prot  \to \mathrm{D} +\ele^+ +\nu_\ele
\label{eq:defpp}
\end{equation}
%It is at the base of the production of $^4$He, emitting 26.7 MeV in the form of photons.
\corrca{This reaction is slow and occurs overcoming the electromagnetic repulsion of the protons, thanks to 
quantum tunnelling: this is why} stars can live for billions of years.

The {\em solar neutrinos,} i.e., the neutrinos emitted by the Sun, are of physical interest for the many reasons:
\begin{enumerate}
\item they tell us how the Sun works, which is important on its own and also for the stellar physics at large;
\item neutrinos escape from the Sun in about 2 seconds, rather than some $100,000$ years as the photons do: 
they are real-time messengers from the Sun;
\item the measurements of solar-neutrino flavor-oscillations give  \corrca{clear evidence of physics beyond the standard model of particle physics.}
\end{enumerate}
The neutrinos produced by the reaction in Eq.~(\ref{eq:defpp}) are the most abundant ones and are 
called pp neutrinos.

\subsection{The Standard Solar Model and its tests}
An essential contribution to the study of solar neutrinos - e.g., \cite{bahcall3,bahcall1,bahcall2} - is due to John Bahcall, a nuclear physicist who came up with what nowadays is called the Standard Solar Model (SSM). 
\corrca{Very roughly, the SSM consists in a physical description of the composition of the Sun and of the processes which make it work.}

Using such model, Bahcall could predict the rate of neutrinos from the Sun and thus \corrca{set} a benchmark for the Homestake experiment (1967-1995), which resulted in the first solar neutrino detection for which Raymond Davis Jr.~was awarded a Nobel prize \cite{homestake}.
However, the number of neutrinos measured \corrca{by} Davis' experiment resulted to be only 1/3 of that predicted by Bahcall, so that both physicists felt compelled to check their procedures.
Their goal was indeed very tough\footnote{We encourage reading the words of Bahcall himself at the following link: \url{http://www.sns.ias.edu/~jnb/Papers/Preprints/Neutrino2002/paper.pdf}.} (and even today solar neutrino detection and modelling is far from easy): on the one hand the SSM is based on models of the metallicity, opacity and of nuclear processes of the Sun (and tested by helioseismology),  
but on the other hand neutrinos interact so weakly that their detection rate is 
small and a very high radiopurity (knowledge and reduction of the sources of background) is mandatory.

Today we know that the Homestake experiment was right, as it was 
sensitive to one neutrino flavor, $\nu_\ele$, being based on the electron neutrino capture process:
\begin{equation}
\nu_\ele + \mathrm{\textsuperscript{37}Cl} \to \mathrm{\textsuperscript{37}Ar}+\ele^-
\end{equation}
Instead, newer experiments as 
Borexino, SNO and Super-Kamiokande can (could) detect neutrinos via their elastic scattering off electrons:
\begin{equation} \label{es}
\nu + \ele^- \to \nu + \ele^- 
\end{equation}
and thus are (were) sensitive to all flavors of (anti)neu\-trinos. This is  
why we used the generic notation $\nu$ for neutrinos here above.
This is even more true for the most important measurement of the SNO experiment,
that was equally sensitive to all three types of neutrinos, and that has been recognized by the Nobel award. 

To give an idea of how complex and interesting the physics of \corrca{the} Sun is, new and more refined versions of the Standard Solar Models are put forward almost every year. The last one 
is  \cite{latestsolarmodel}, 
published in \corrca{2017}.

\begin{center}
\begin{figure*}\centering
\begin{minipage}[c]{11.3cm}
\includegraphics[width=1\textwidth,keepaspectratio]{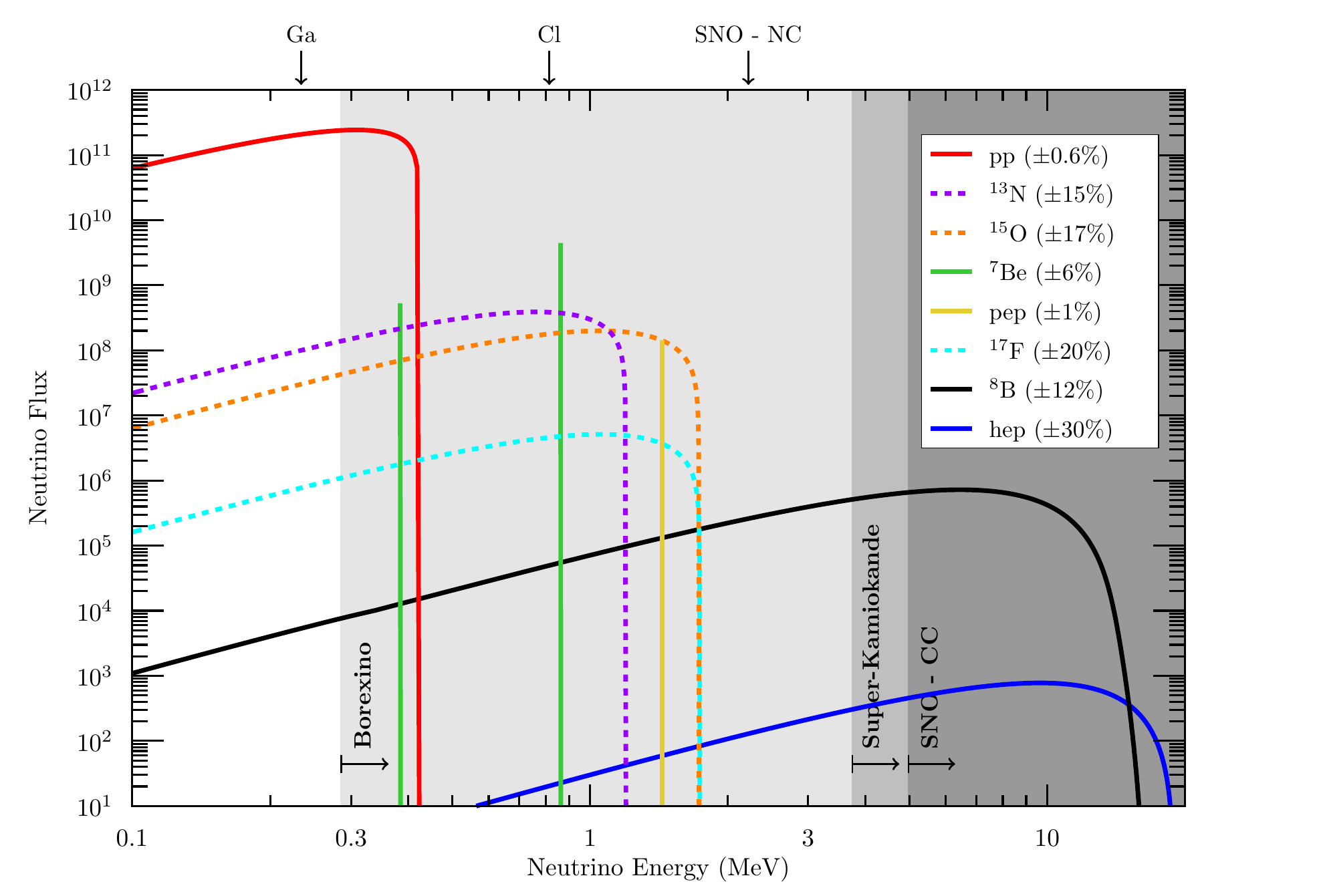}
	\end{minipage}
	\begin{minipage}[c]{6.0cm}
\caption{Energy dependence of the neutrino fluxes produced by the different nuclear processes in the Sun, according to the B16-GS98 
Standard Solar Model \cite{latestsolarmodel}. 
The $y$ axis is in units of cm$^{-2}$ s$^{-1}$ MeV$^{-1}$ for continuous spectra and cm$^{-2}$ s$^{-1}$ for monochromatic ones (beryllium and $\prot\ele\prot$ neutrinos).
The neutrino energy threshold for detection is indicated with a vertical arrow for the experiments which do not reconstruct the neutrino energy (those based on gallium, as GALLEX/GNO and SAGE, on chlorine, as Homestake, or 
on neutral current $\nu$-D interactions, as SNO) and with \corrca{a} horizontal arrow for those which reconstruct the neutrino energy (those based on elastic scattering, as Borexino and Super-Kamiokande, and on charged current $\nu$-D interactions, as SNO).
The dashed lines highlight the CNO\corrca{-cycle} contribution to the solar neutrino flux.}
\label{solarnu}
	\end{minipage}
\end{figure*}
\end{center}

\subsection{The pp neutrino flux and their detection}
The nuclear reaction in Eq.~\eqref{eq:defpp} begins a set of reaction called  $\prot\prot$ chain: the SSM predicts that the $\prot\prot$ chain is the main channel of energy and neutrino production, as reported in Table \ref{tab:2} and as evident from figure \ref{solarnu}.
\begin{table}[t]
	\caption{\small{\em SSM expectations of the solar neutrinos arriving at Earth, 
	due to the various branches of neutrino production \cite{latestsolarmodel}. The first 5 neutrinos
	are those of pp-chain (all observed except the last); the last 3 are those of the CNO-cycle (yet unobserved).}}
     \vskip 2 mm
\begin{center}
\begin{tabular}{clrr} \normalsize
branch &&flux, $10^6/$(cm$^{2}$ s) &$E_\nu^\text{\tiny max}$, MeV\\
\midrule \midrule
$\prot\prot$ 	&&$59,800	(1\pm  0.006) $		&0.420\\ \midrule
$^7$Be 	&&$4,930 	(1\pm    0.06) 	 $ &0.862 \\ \midrule
$\prot\ele\prot$  	&&$144	(1\pm    0.01) 	 $ &1.442 \\ \midrule
$^8$B 	&&$5.46	(1\pm    0.12) 	$  &15.1$\phantom{00}$  \\ \midrule
$\hh\ele\prot$  	&&$0.008	(1\pm    0.30)	 $ &18.773\\ \midrule \midrule
$^{13}$N 	&&$278	(1\pm    0.15) $	  &1.199	\\ \midrule
$^{15}$O 	&&$205	(1\pm    0.17) $	 &1.732 \\ \midrule
$^{17}$F 	&&$5.28	(1\pm    0.20) $ &1.740   \\ \midrule \midrule
\end{tabular}
\end{center} 
\label{tab:2}
\end{table}
The $\prot\prot$ neutrinos amount to more than 90\% of the solar neutrino flux coming \corrca{to} Earth.

The flux can be 
roughly estimated by elementary considerations: 
We know that a nucleus of Helium is formed after two $\prot\prot$ fusions, so that:
\begin{equation} 
\dot N_{\prot\prot} \simeq 2 \dot N_{\mathrm{He}} 
\end{equation}
Moreover, we know that $Q=26.7\;$MeV are produced in the production of Helium, which implies
\begin{equation}
\dot N_{\mathrm{He}} \simeq \frac{L_\odot}{Q} \simeq 10^{38} \,\mathrm{Hz} \end{equation}
In this manner, we obtain an estimate of the flux, 
\begin{equation}
\Phi_{\nu_\ele} \simeq \frac{2 \dot N_{\mathrm{He}}}{4\pi (1\;\mathrm{AU})^2}\simeq 6 \times 10^{10} \;\mathrm{cm}^{-2}\,\mathrm{s}^{-1}
\end{equation}
This flux is sizeable, but, considering experiments which detect neutrinos via elastic scattering of Eq.~(\ref{es}), we need to keep in mind that the probability of interaction (cross-section) of neutrinos with electrons is extremely small, even though theoretically very clean: for the cases of our concern, it grows \agr{with the square of the neutrino energy} $E_\nu^2$.
This means that, despite such a huge flux of $\prot\prot$ neutrinos, it is very difficult to detect them, and in fact their detection is recent \cite{borexpp}.
Moreover, neutrino-electron scattering detection has a threshold (on the recoil kinetic energy of the scattered electron) as low as 150 keV (in Borexino) with a neutrino energy reconstruction accuracy of the order of few percent.

Neutrino detection in Borexino is in fact based on the scintillation light emitted by the electron on which the neutrino impinged; Cherenkov detection has been discarded due to its very small light yield compared to that of scintillation \corrca{(a few \%)}.
Since the scintillation light is emitted isotropically, direction reconstruction is not feasible / is very very difficult 
in this type of detectors\footnote{\rvw{As will be discussed later, this is possible in other types of detector, such as the Super-Kamiokande. Note, however,  
that the electron tracks the direction of the neutrino only if $E_\nu\gg m_\ele$.}}, \rvw{even} in the case of elastic scattering Eq.~(\ref{es}).

Moreover, neutrino-induced events in liquid scintillators are intrinsically indistinguishable on an event-by-event basis from the background due to $\beta$ or $\gamma$ decays: this is why radiopurity is a key factor of this kind of experiments, aimed at finding rare neutrino events at very low energies.

\begin{figure*}
 \begin{center}
 \begin{minipage}[c]{8.5cm}
\includegraphics[width=0.9\textwidth,keepaspectratio]{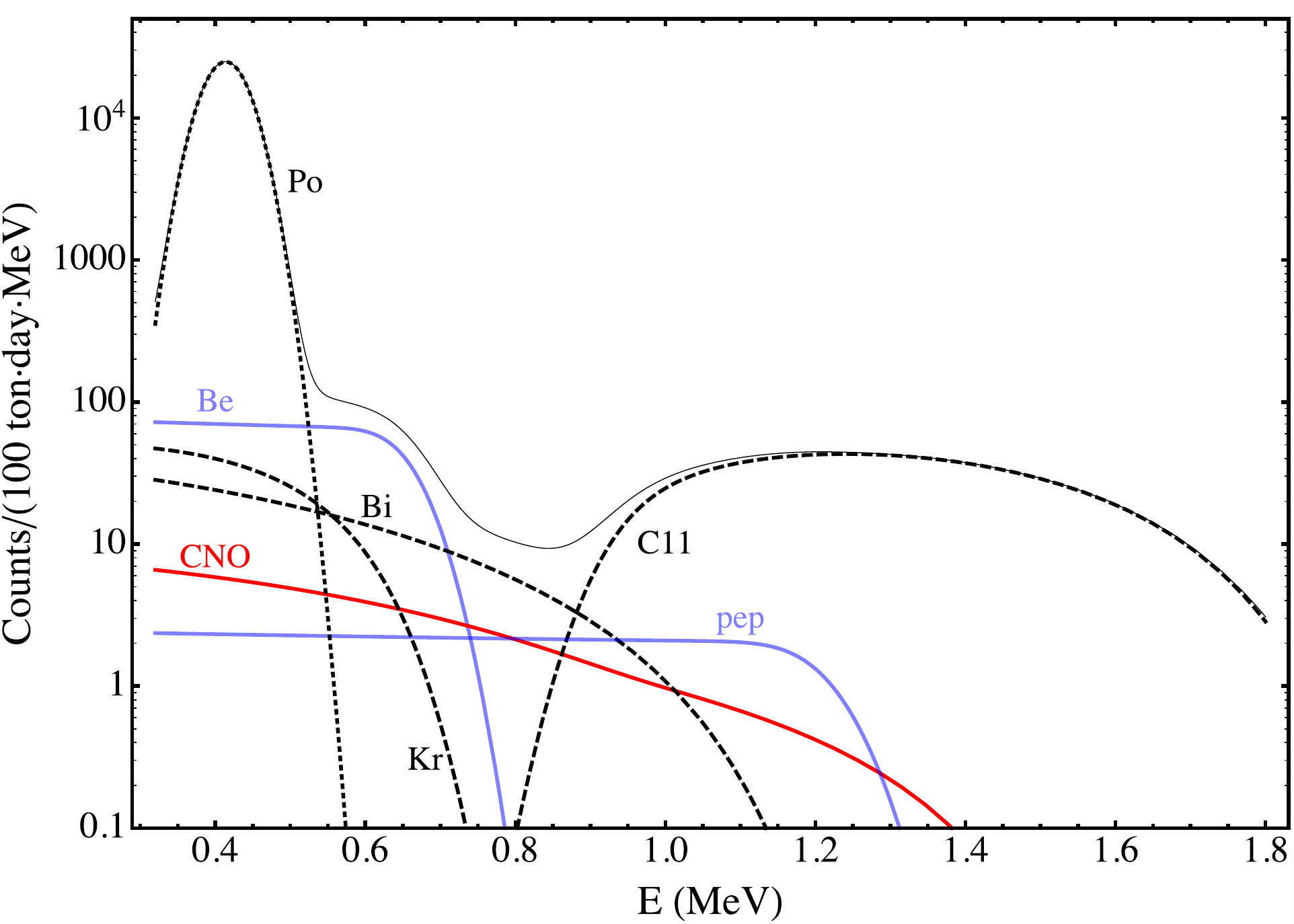}
\end{minipage}
\hfill
	\begin{minipage}[c]{8.1cm}
	\caption{The differential rate of events in Borexino - inclusive of the background - as a function of the electron recoil kinetic energy. The continuous lines are the signals due to solar neutrinos; the CNO neutrinos indicated by a red line, 
the neutrinos from the PP chain are indicated by blue lines. The dotted and dashed black lines are due to background events instead. Particularly evident is the importance of a good knowledge of the bismuth background in order to detect CNO neutrinos.
Figure adapted from  \cite{vborex}.}
\label{cno} 
	\end{minipage}
	\end{center}
\end{figure*}

\subsection{What we (do not) know}
Once again, the state of the art \corrca{of} solar neutrino expectations is summarized in table \ref{tab:2} and figure \ref{solarnu}.

For what concerns the measurements, we emphasize that the reconstruction of the energy of the solar neutrinos is important to test our understanding of  
how the Sun works. Super-Kamiokande and SNO measured accurately the ${}^8$B neutrinos, which are relatively more easy to be observed
but are a very small \corrca{fraction} ($\sim$0.02\%) of the SSM neutrino flux.
Borexino experiment observed directly pp, pep, and ${}^7$Be neutrinos, measuring the latter precisely.
We have only upper bounds on the very rare hep neutrinos--see again table~\ref{tab:2}.

In short, most of the neutrinos from the pp-chain have been measured, while those from 
the CNO cycle are not. The CNO cycle in the Sun is expected to yield $\sim 1$\% of the solar luminosity,
but it is very important for astronomy, being the main cycle of energy production in massive stars. 
It begins with a set of reactions called CNO-I cycle, 
\begin{align*} 
\phantom{centramiiiiiiii}\phantom.^{12}\mathrm C + \prot &\to \phantom.^{13}\mathrm N+\gamma\\
\phantom.^{13}\mathrm N &\to \phantom.^{13}\mathrm C +\ele^++\nu_\ele\\
\phantom.^{13}\mathrm C+\prot&\to \phantom.^{14}\mathrm N +\gamma\\
\phantom.^{14}\mathrm N+\prot&\to \phantom.^{15}\mathrm O+\gamma\\
\phantom.^{15}\mathrm O &\to \phantom.^{15}\mathrm N +\ele^++\nu_\ele\\
\phantom.^{15}\mathrm N + \prot &\to \phantom.^{12}\mathrm C+\phantom.^4\mathrm{He}
\end{align*}
As we can see, CNO-I yields 
neutrinos from the $\beta^+$ decay of ${}^{13}\mathrm N$ and ${}^{15}\mathrm O$.
Replacing  the last reaction of 
$\alpha$ emission with $\gamma$ emission we begin the CNO-II cycle, that is much less frequent in our Sun,
%about 0.04\% of the times,
\begin{align*} 
\phantom{centramiiiiiiii}
\phantom.^{15}\mathrm N + \prot &\to \phantom.^{16}\mathrm O+\gamma\\
\phantom.^{16}\mathrm O+\prot &\to \phantom.^{17}\mathrm F +\gamma \\
\phantom.^{17}\mathrm F&\to \phantom.^{17}\mathrm O+\ele^+ +\nu_\ele\\
\phantom.^{17}\mathrm O+\prot&\to \phantom.^{14}\mathrm N+\phantom.^4\mathrm{He}
%\phantom.^{14}\mathrm N +\prot&\to \phantom.^{15}\mathrm O+\gamma\\
%\phantom.^{15}\mathrm O + \prot &\to \phantom.^{15}\mathrm N+\ele^++\nu_\ele
\end{align*}
Also this cycle yields one type of neutrinos, due to $\beta^+$ decay of ${}^{17}\mathrm F$. 
Finally, there are also CNO-III and CNO-IV cycles,  important only for very massive stars.

CNO neutrinos offer us a chance to understand the CNO cycle and Borexino has a chance to see them for the first time; 
unfortunately, this measurement represents a true challenge, as can be understood from figure \ref{cno}.
The $\beta$ decay of bismuth (one of the daughters of radon) is the main reason of background events. However, 
% it can be measured by the 
it is associated to the $\alpha$ decay of polonium \cite{vborex}, that leads to events 
visible on the left hand side of  figure~\ref{cno}
\begin{equation}
{}^{210}\mbox{Bi} \stackrel{\mbox{\tiny 7d}}{\longrightarrow} 
{}^{210}\mbox{Po} +\mbox{e}^-+\bar\nu_{\mbox{\tiny e}} \stackrel{\mbox{\tiny 200d}}{\longrightarrow}
{}^{206}\mbox{Pb} +\alpha
\end{equation}
%\begin{align*} 
%\phantom{centra}
%\phantom.^{210}\mathrm{Bi} &\to \phantom.^{210}\mathrm{Po}+\overline \nu_\ele + \ele^- & \tau =7.2\mbox{ days}\\  \phantom.^{210}\mathrm{Po}&\to \phantom.^{206}\mathrm{Pb}+\alpha & \tau=200\mbox{ days}\end{align*}
In stable conditions, the decay events follow the law of radioactive decay which 
should allow to measure the bismuth from polonium \cite{vborex}.  
Borexino has the chance to measure CNO neutrinos in a few years 
applying a dedicated analysis.

%\clearpage
\section{Supernova neutrinos\label{sec:sn}}

\corrca{Neutrinos emitted during the gravitational collapse, associated to a visible supernova of a certain astronomical type, are} called for short  {\em supernova neutrinos}. 
The astrophysics is much more complicated than \corrca{that} of solar neutrino emission, the associated uncertainties are large and the theorists involved in the study of this type of events are still unsure whether their computer simulations include all the relevant physics. 

However, 30 years ago we had a successful observation of the neutrinos from one supernova, called SN 1987 A, that has been recognized by the 2002 Nobel prize in physics to the Koshiba, the leader of \corrca{the} \agr{Kamio\-kande} experiment. 

In order to provide an \corrca{introduction to} this interesting and complex subject, we will focus the discussion on basic features of neutron stars and of the gravitational collapse; on expectations about the neutrino emission; on the signal of electron antineutrinos in \rvw{detectors devoted to monitor supernova emission}, with a brief mention of the events from SN1987A. 
\rvw{For further general discussion, see \cite{astro-book,rev1}; for a specialized discussion 
on SN1987A, see \cite{Vissani:2014doa}.}

\subsection{The supernova - neutron star connection}
Supernovae have been occasionally observed in the past centuries. 
In 1933, just after the discovery of the neutron, Baade and Zwicky proposed the idea that this observable astronomical phenomenon is connected with the formation of a very compact object: a large fraction of the stellar mass gets transformed from ordinary atomic matter into neutrons. 
In other words, the size, typically occupied by matter, decreases from the one corresponding to atoms, $R_{\mbox{\tiny atomic}}\sim 10^{-8}$ cm to the one corresponding to nuclei, $R_{\mbox{\tiny nucl}}\sim 10^{-13}$ $\mathrm{cm}=1$ fermi, namely, it would decrease by $R_{\mbox{\tiny atomic}}/R_{\mbox{\tiny nucl}}\sim 10^5$ times. 
\agr{For example}, our Sun would fit in a region of $700\,000$ km/$10^5\sim$ 7 km! 

Indeed, the typical radius of a neutron star (estimated by using general relativity and the theory of strong interactions) is $R_{\mathrm{ns}}=10-15$ km. 
\corrca{On the other hand}, its typical (measured) mass is $M_{\mathrm{ns}}\ge 1.44\, M_\odot$ or larger, which is the maximum mass of a white dwarf before that it collapses under its weight. 
The existence of such a limiting mass\footnote{Its value is $\sim M_{\mathrm{Pl}}^3/m_\neut^2$, where $m_\neut=1/(6\times 10^{23})$~g is the neutron mass and $M_{\mathrm{Pl}}=\sqrt{\hbar c/G_{\mathrm{N}}}=2\times 10^{-5}$~g the Planck~mass.}  
was predicted in 1930 by Chandrasekhar.

The \agr{gravitational binding energy $\mathcal{E}$}
of the neutron star is huge:
\begin{equation} \label{bindament}
\mathcal{E}\sim \frac{3}{5} \frac{G_{\mathrm{N}} M_{\mathrm{ns}}^2}{R_{\mathrm{ns}}}=(2-3)\times 10^{53}\,\mathrm{erg}
\end{equation}
where of course $G_{\mathrm{N}}\sim 7\times 10^{-8}$ erg cm/g$^2$ is the Newton constant.
This is a significant \corrca{(10-15\%)} fraction of the rest mass $M_{\mathrm{ns}} c^2$; in order to allow the formation of the neutron star, this energy should be released. 
The main agents which \corrca{are able to carry away such tremendous amount of energy in a short time are} just neutrinos. 
In fact, these particles are copiously produced during and immediately after the collapse of the star, and can escape from the star rapidly enough. 

In order to better \corrca{picture} how this happens, consider the opacity of neutrinos crossing the neutron star:
\begin{equation}
\tau=\rho/m_N\times \sigma\times R_{\mathrm{ns}}
\end{equation}
With a cross section of $\sigma\sim 10^{-41}$ cm$^2$, with 
a density of $\rho=10^{11}$ g/cm$^3$, with a radius of 10 km, we have $\tau\sim 10^{11}\times 6\cdot 10^{23}\times 10^{-41} \times 10^6\sim 1$, which means that the external regions of the neutron star are transparent and allow neutrinos to \corrca{escape}. 
\rvw{In the innermost parts of the neutron star, however, we have $\sim 1$ particle/fermi$^3$. Thus 
the density is $\rho\sim$few$\times 10^{14}$ g/cm$^3$, and neutrinos are trapped in the region where they are produced.}
The cross section $\sigma\sim 10^{-41}$ cm$^2$ 
corresponds to neutrinos with energies\footnote{This typical energy can be considered just in between two extreme (and opposite) cases for neutrino radiation. The first case corresponds to a star that is {\em as opaque as possible} and radiates neutrinos only from the center; we would have $\tau\sim 1$  for  $\rho\sim$ 1 particle/fermi$^3$ and $\sigma\sim 10^{-44}$ cm$^2$, which means very low energy (1 MeV) neutrinos. The second case corresponds to a {\em transparent} star, where any electron would become an electronic neutrino, with energy of few times 100 MeV.} 
of $10-20$ MeV. The surface of the star, where neutrinos are radiated, is called neutrinosphere and can be thought of, in a first approximation, as a black body radiator. 

The sequence of the events, which leads to the formation of a neutron star, is as follows; 
\begin{enumerate}
\item in the largest stars (above 6-10 $M_\odot$) a chain of nuclear reactions  
leads to form a core of iron, that is stable \corrca{with respect to} further reactions;
\item when the core reaches the Chandrasekhar mass, the electrons become relativistic and the associated pressure is unable to support the core, that collapses under its weight;
\item the radius becomes 10 km, and the pressure  of the neutrons provides stability to the new configuration;
\item the huge potential energy is radiated in neutrinos of all types, with energies around 10-20 MeV.
\end{enumerate}
A small fraction of this energy, much less than Eq.~(\ref{bindament}),  
is converted then into kinetic energy. 
Indeed, the masses of gas expelled by a supernova are observed to have kinetic energies of the order of $M v^2/2=10^{51}$ erg, which corresponds to a mass of $M\sim 10\ M_\odot$ moving at a velocity of $v\sim 3,000$ km/s.
However the details of how this conversion takes place are not reliably known yet.

% arrivare alla stima della energia totale e della energia media dei neutrini emessi
\begin{figure*}
    \begin{center}
	\begin{minipage}[c]{8.5cm}
    \caption{Expected distribution ($=$ spectrum) of the IBD signal due to electron antineutrinos, for a future galactic supernova exploding at 10 kpc and for a neutrino detector comprising $2.1\times 10^{33}$ protons--e.g., 32 kton of water or 29 kton of linear alkyl benzene.  
    The expectations and the uncertainties are modeled on SN1987A data, see  \cite{astro-book,Vissani:2014doa} for details.}
      \label{fig:rangion}
    \end{minipage}
    \hfill
 \begin{minipage}[c]{8.cm}
\includegraphics[width=0.9\textwidth]{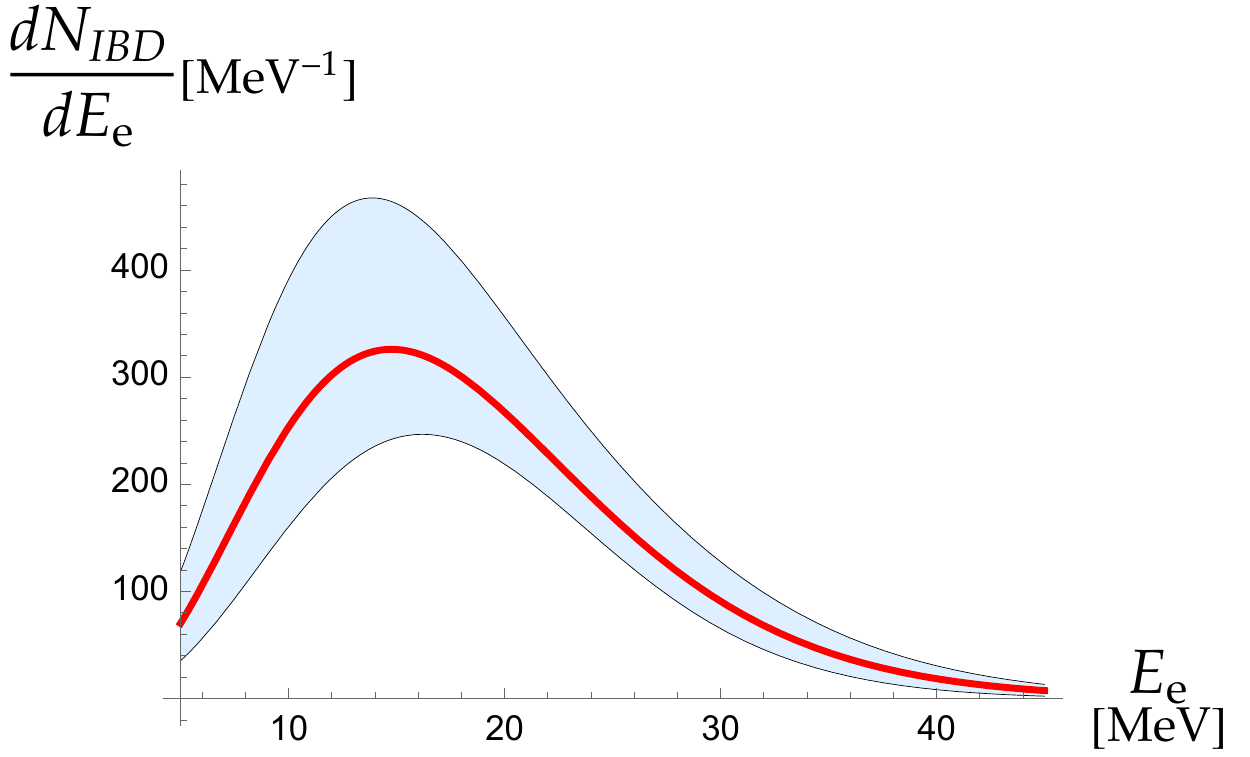}
\end{minipage}
    \end{center}
\end{figure*}

\subsection{The expected neutrinos}
Let us begin by estimating the time scale of emission. 
The 6 types of neutrinos and antineutrinos are supposed to carry away the surplus of energy, thereby allowing the formation of the neutron star. 
Let us assume that (most of) them are emitted close to the surface of the proto-neutron star simply as black body radiation, and moreover that the temperature of each type of (anti)neutrinos is close to $T_\nu \sim 4$ MeV.
The luminosity (=the radiated power) \corrca{per neutrino type} can be directly estimated from the Stefan-Boltzmann law: 
\begin{equation} 
L_\nu\approx R_{\mathrm{pns}}^2\times T_\nu^4 
\mbox{ with }\nu=\nu_{\mbox{\tiny e}},\nu_\mu,\nu_\tau,\bar{\nu}_{\mbox{\tiny e}},\bar\nu_\mu,\bar\nu_\tau
\end{equation}
where the radius of the proto-neutron star is around $R_{\mathrm{pns}}\sim 15$ km. 
By including the dimensional constants $c=3\times 10^{10}$ cm/s and $\hbar c=200$ MeV$\times$fermi, the value of the luminosity in 
conventional units turns out to be:  
\begin{equation} 
L_{\mathrm{tot}}=6\times L_\nu \approx 2\times 10^{52}\text{erg s}^{-1} 
\end{equation}
Thus, the whole potential energy of the star is gone in a time $\mathcal{E}/L_{\mathrm{tot}}$, 
namely, 10 seconds or so.

Next, we would like to estimate the \corrca{neutrino} fluence on Earth,
i.e.\ the flux integrated over the time of emission. 
Let us begin by calculating the number of neutrinos, that for each species 
($\nu_\ele,\nu_\mu,\nu_\tau,\bar\nu_\ele,\bar\nu_\mu,\bar\nu_\tau$) is
\begin{equation}
N_\nu =\frac{\mathcal{E}}{6 \langle E_\nu \rangle} \approx 
\frac{3\times 10^{53}}{6\times 10 } \frac{\mathrm{erg}}{\mathrm{MeV}} \approx \mathrm{few}\times 10^{57} \frac{\mathrm{erg}}{\mathrm{MeV}}
\end{equation}
For comparison, we note that the number of electrons in the iron core (i.e., the electronic lepton number of the collapsing object) is,
\begin{equation}
N_{\ele}=\frac{Y_\ele\; M_{\mathrm{core}}}{m_N}=0.4\times 3\times 10^{33}\times 6\times 10^{23}=7\times 10^{56}
\end{equation}
namely, it is more than one order of magnitude smaller than the total number of emitted neutrinos. 
Using \corrca{as benchmark} a distance of $D=10$ kpc, which is a typical galactic scale, we find that the neutrinos distribute \corrca{o}n the surface of a sphere with area $S=4\pi D^2=10^{46}$ cm$^2$ before reaching the Earth, and therefore, we have, 
\begin{equation}
F_\nu=\frac{N_\nu}{S}=\mathrm{few}\times 10^{11} \frac{\nu}{\mathrm{cm}^2}
\end{equation}
where we recall that the time-integrated flux $F_\nu=\int \Phi_\nu(t) dt$ is called {\em the fluence}.

\subsection{The IBD signal, SN1987A and the future}
The type of neutrinos that has the largest probability of interaction with ordinary matter are the electron antineutrinos. 
The relevant reaction is in fact the one that has been first considered for the detection, that is sometimes called ``inverse beta decay'' (IBD), namely:
\begin{equation}
\overline\nu_\ele+ \prot\to \neut + \ele^+
\label{ibd}
\end{equation}
where the interaction happens with a proton, i.e., a nucleus of hydrogen.\footnote{Let us recall that this is 
abundant in water or in hydrocarbon compounds, the most common type of materials used for supernova neutrino detection.}
A typical value of the cross section at the relevant energies is $\sigma_{\mbox{\tiny IBD}}\sim 1.5 \times 10^{-41}$ cm$^2$. 
If we consider the detector Super-Kamiokande \cite{sk}, that has 22.5 kton of water, we have that the number of target protons is $N_\prot=1.5\times 10^{33}$. 
Therefore, using the fluence $F_{\overline\nu_\ele}=2\times 10^{11}\ \overline\nu_\ele/$cm$^2$ estimated above for a supernova at 10 kpc, we find that the expected number of positrons $\ele^+$ due to the IBD reaction is: 
\begin{equation}
N_{\ele^+}= N_\prot\times \sigma_{\mbox{\tiny IBD}}\times F_{\bar\nu_\ele}\approx 4,500
\end{equation}
which is about 200 times \corrca{larger than what} was seen in the case of SN1987A.\footnote{Indeed, the distance of SN1987A  (52 kpc) was  
5 times larger and the mass of Kamiokande (2.14 kton) was 
10 times \corrca{smaller}; then the approximate scaling factor is $5^2\times 10=250$.}
For an accurate prediction including uncertainties see Fig.~\ref{fig:rangion}, adapted from ref.~\cite{astro-book}.

Before \corrca{leaving this topic}, it should be recalled that the number of events seen from SN1987A in the same time window was small: 12 in Kamiokande, 8 in IMB, 5 in Baksan. 
In all cases, the emission lasts about 10 seconds and almost half of the IBD events (6, 3 and 2 respectively) happened in first second: this is consistent with the simulations, and will be one of the most interesting things to test with a future event. 

Another very important thing that we hope to measure in the future \rvw{with Super-Kamiokande \cite{sk}}
is the sample of events from the neutrino scattering \corrca{off} atomic electrons 
%\begin{equation}
introduced in Eq.~(\ref{es}) for solar neutrinos.
%\end{equation}
This is a directional reaction, i.e., the electrons are emitted from the direction of arrival of the neutrinos \rvw{within some tens of degree for each individual event. This will allow a few degrees reconstruction of the supernova position, see e.g., \cite{astro-book}.} Moreover, 
this reaction is not exclusively sensitive to $\overline\nu_\ele$ (as in the case of IBD)
but also to 
$\nu_\ele,\nu_\mu,\nu_\tau,\overline\nu_\mu,\overline\nu_\tau$.

\section{High energy cosmic neutrinos\label{sec:cn}}
\corr{Before discussing high energy cosmic neutrinos we have to know what are the different backgrounds that can make their detection} \corrca{difficult} (sec.~\ref{sec:CRbkg}). Then we discuss theoretical expectations on the cosmic neutrino flux (sec.\ \ref{sec:CRsignal}). We discuss the possible signals in neutrino \rvw{detector and telescopes} and the observational opportunities (secs.\ \ref{sec:CRgammanu3}-\ref{sec:CRgammanu2}).
Finally, we  examine in the appendix the connection between neutrino and ordinary astronomy.
\rvw{For a few general references, see \cite{astro-book,baur,mau,ger}; to go deeper into the subject, 
see the specialized review works \cite{a5,a1,a2,a3,a4,a6,a7,a8}.}

\subsection{High energy neutrinos of atmospheric origin}
\label{sec:CRbkg}
Neutrinos of high energy are produced in our atmosphere. In fact, {\em atmospheric neutrinos} come from the decay of unstable particles (mesons) produced by the
cosmic ray bombardment that hits our atmosphere; these unstable particles have an energy that extends on average from 100 MeV \corr{up to} many TeV.  The atmosphere acts initially as a converter, that degrades the energy of the primary nucleons $N_{\mbox{\tiny cr}}$ 
producing mesons and eventually secondary particles
\begin{equation}
N_{\mathrm{cr}} + N_{\mathrm{atm}} \to X+ \text{\small many mesons}\left\{ 
\begin{array}{l} \pi^0\to \gamma+\gamma \\[1ex] 
\pi^\pm\to \mu^\pm + \oset[0.35ex]{\mysymb}{\nu}_\mu 
\label{e1}
\end{array}
\right.
\end{equation} 
The $\gamma$-rays of high energy are screened by the atmosphere itself,  
while neutrinos and \corrca{high-energy} muons reach the ground and are observable.
The path length of the muon is $\gamma_\mu \tau_\mu c=(E_\mu/m_\mu)\times 0.6 $~km, therefore when their energy
is above 10 GeV, muons reach the ground before decaying, and the neutrinos that we receive on Earth 
are just the $\oset[0.35ex]{\mysymb}{\nu}_\mu$ of Eq.~(\ref{e1}).

The neutrino spectrum behaves as $\sim E_\nu^{-2.7}$ at low energies,  \corr{up to} few GeV, just as the primary cosmic rays;\footnote{This is a property of the collisions of cosmic rays and hadrons, that is called {\em scaling}.} 
then it 
becomes steeper for the following reason.
%The reason is as follows.
Charged pions with more than 10 GeV have a Lorentz factor of 
$\gamma_\pi> 70$ and travel  $\gamma_\pi \tau_\pi  c> 0.4$ km. Now, let us calculate the 
interaction length. For an 
atmospheric density of about \corr{$\rho=10^{-3}$ g/cm$^3$} and for 
a nuclear interaction cross section  $\sigma_n=\pi r_n^2$ 
corresponding to a radius of $r_n=1$ fm, 
we have a mean free path of 
$\lambda_n=1/(N_A\rho\sigma_n)=0.5$ km, \corr{where $N_A$ is the Avogadro number}. Therefore, the higher the pion energy, 
the smaller their chance to decay before \corrca{losing} a significant fraction of energy. 
% 
%
%that the high energy pions interact with the atmosphere: In fact, for an 
%atmospheric density of about $\rho=$1 g/cm$^3$ and for 
%a nuclear interaction cross section  $\sigma_n=\pi r_n^2$ 
%corresponding to a radius of $r_n=1$ fm, 
%we have a mean free path of 
%$\lambda_n=1/(N_A\rho\sigma_n)=1/2$ km, which is much less than the size of the atmosphere. 
%Charged pions with more than 10 GeV have a Lorentz factor of 
%$\gamma_\pi> 70$ and travel  $\gamma_\pi \tau_\pi  c> 0.4$ km: this means that they interact 
%before decaying and loose energy, just as the primary cosmic rays. 
For this reason, 
atmospheric neutrinos are distributed as $\sim E_\nu^{-3.7}$ in the region 100 GeV-100 TeV and 
depleted at higher energies.

%\begin{center}
\begin{figure}[t]
\centerline{\includegraphics[width=0.34\textwidth,keepaspectratio]{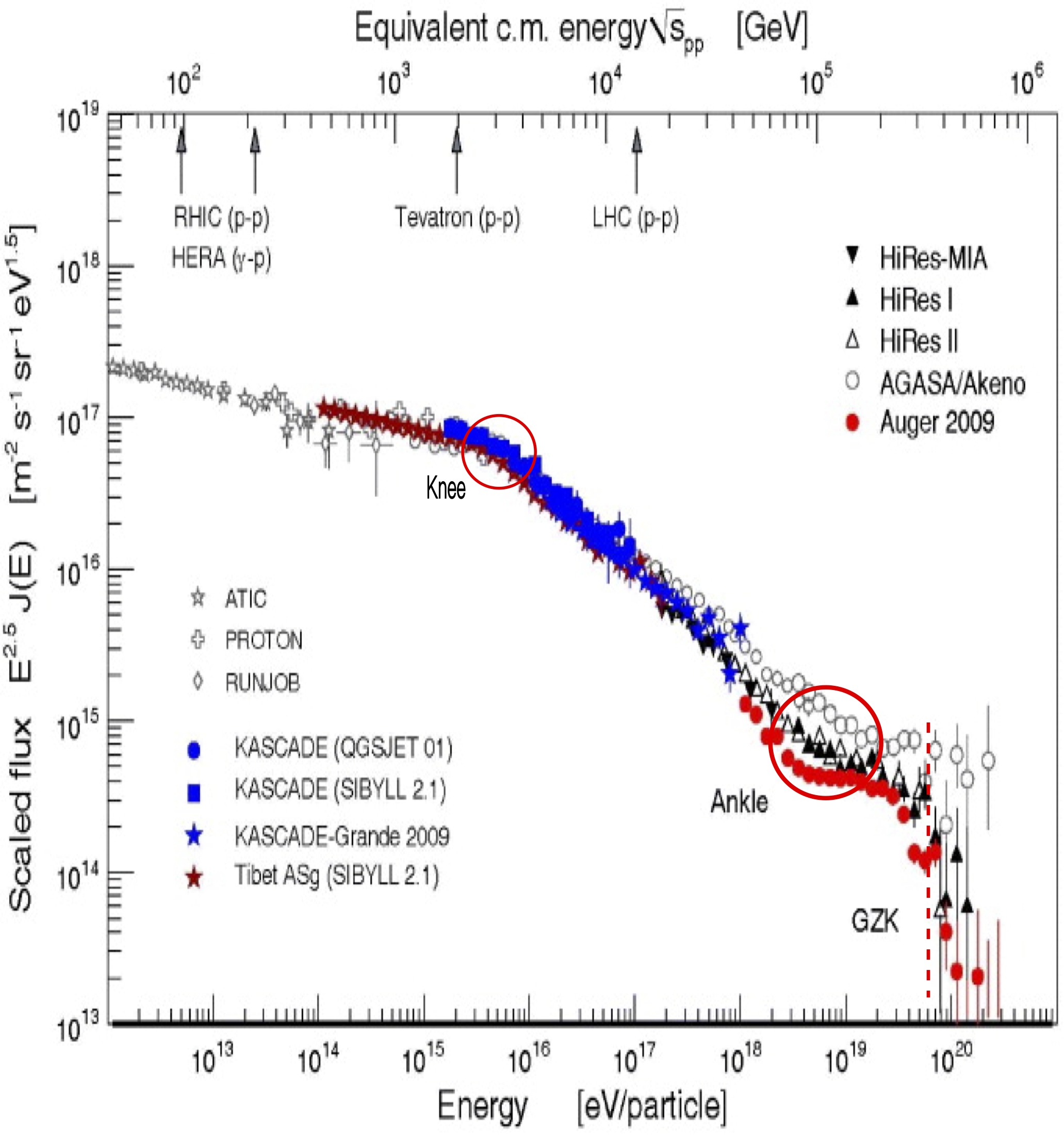}}
\caption{The cosmic ray spectrum, slightly modified version of figure 2 of \cite{matthiae}.}
\label{fig:cr_spectrum}
\end{figure}
%\end{center}

The \corrca{most energetic pions carry} away $\sim 1/5$ of the \corrca{energy of the primary}, then \corrca{every} decay particle carries away\corrca{, on average,} a similar amount of energy.
There are two $\gamma$-rays for the $\pi^0$ decay and four particles
for the full decay of the $\pi^\pm$, since 
$\mu^\pm \to \ele^{\pm} +\oset[0.35ex]{\mysymb}{\nu}_\ele+\oset[0.35ex]{\mysymb}{\nu}_\mu$
so, on average, 
the photons are \corrca{twice as energetic as} the corresponding neutrinos. Thus, each $\gamma$-ray carries away about 1/10 of the primary energy, while this fraction becomes 1/20 for the neutrino.

Recalling that the cosmic ray spectrum shown in figure \ref{fig:cr_spectrum} at Earth softens above the 
{\em knee}, with $E_{\mathrm{knee}}=$ few PeV, we expect that the atmospheric neutrino spectrum 
extends \corr{up to} 
\begin{equation} \label{genuc}
\frac{E_{\mathrm{knee}}}{20}\sim 100\mbox{ TeV}
\end{equation}  
\corrca{For higher energies}, the sky should become more and more 
free from atmospheric background events; thus, the high energy neutrinos could reveal the 
sources of the cosmic rays. The region below few 100 TeV is polluted by atmospheric neutrinos (and muons)\corrca, but it is 
also important -- in particular for the search of the sources of galactic cosmic rays, that are  usually 
believed to have energies lower than  the knee.

\corr{Not only pions and kaons, but also} charmed particles are produced in pairs in cosmic ray collisions. 
Even if they are less common than pions, their lifetime is so short that they immediately decay and yield a spectrum of {\em prompt} neutrinos with shape 
$\sim E_\nu^{-2.7}$ that reflect the cosmic rays that impinge on atmosphere. Thus, somewhere at high energies the contribution of this component should exceed the one due to pions and kaons; this should happen around 100 TeV, even if the theoretical estimations are not stable enough to provide a solid prediction.
Two main search strategies are possible: \begin{enumerate} 
\item the prompt neutrino spectrum (whose shape is assumed to be known) is fitted to the muon neutrino flux in order to extract the normalization of the prompt neutrino spectrum - and this is currently implemented in IceCube;
\item the other strategy is based on the study of the electron neutrino component, that is as abundant as \corrca{that of} muon neutrinos \corrca{coming} from the decay of charmed particles, (prompt) whereas it constitutes only 5\% of the conventional \corrca{(from the decay of pions and kaons)} fluxes.   \end{enumerate}
Prompt neutrinos have not been detected yet, and are thought to provide only few events in the detector; the bound currently obtained by IceCube already rules out the most optimistic predictions.
Thus, there are chances to see high energy cosmic neutrinos over the atmospheric ones.

\subsection{Sources of cosmic neutrinos}
\label{sec:CRsignal}

Although we do not have yet a clear theoretical idea on which the sources of cosmic neutrinos are and how intense they are, there are several ideas on the subject. In this section, we examine the case that the cosmic neutrinos are directly connected to the (sources of the) cosmic rays. In the appendix, we elaborate the connection between neutrinos and gamma rays.

\paragraph*{Cosmic rays and cosmic neutrinos:--} % pag.58 delle trasparenze
Consider some astrophysical site where cosmic rays are produced and confined for some time. Suppose that there is some target (say, gas) for the cosmic rays to interact with. Moreover, let us begin by considering the case \corrca{of} this target \corrca{being} much more diffuse than Earth's atmosphere. 
We might imagine that this site will be particularly brilliant in neutrinos and $\gamma$-rays, and that the shape of these secondary particles will reflect the production spectra of cosmic rays. 
The cosmic ray spectrum in the source is unknown {\em a priori}, however, there is one 
general theoretical argument in support of the idea that the spectrum behaves as   
$\sim E_{\mathrm{cr}}^{-2}$, both for $\gamma$-rays and for neutrinos.
1)~First of all, \corr{such a spectrum is expected in the Fermi acceleration mechanism.} 2)~Second, if--just in the terrestrial atmosphere--the 
target for the cosmic ray collisions is composed by protons or other nuclei, but--differently from the atmosphere--the target layer is thin and there is no significant absorption of the mesons--the spectra of the secondary particles at the production site reflect closely the shape of the primary cosmic ray spectrum. 3)~Third, it is not plausible that neutrinos suffer absorption in the source (while for gamma rays, this is possible, as  
discussed in some details in the appendix). In these conditions, the spectrum of the secondary neutrinos 
(and possibly of the associated $\gamma$-rays) till some maximum energy, is expected to be distributed as 
\begin{equation}\frac{d\Phi_\nu}{dE_\nu}\propto E_{\nu}^{-2}\end{equation} 
This spectrum would stand out over the 
atmospheric neutrino background at sufficiently high energies.\footnote{A similar conclusion follows from one  
more precise but even much more specific version of this argument. As we have recalled, the cosmic ray spectrum observed \corrca{at} Earth distribute as 
$\sim E_{\mathrm{cr}}^{-2.7}$ till the knee. This is  thought to be due to \corrca{the} convolution of some (average) production spectrum of galactic sources with diffusion and escape factors, that increase with energy and therefore deplete the population of high energy cosmic rays. Thus, the production spectrum of cosmic rays is expected to be harder. Secondary particles produced by hadronic collisions have the same properties and their spectrum
might extend till $\sim 100$~TeV as in Eq.~(\ref{genuc}).}

\paragraph*{\corr{Extragalactic sources:--}}
%\label{sec:CRextragal}
\agr{Nowadays, extragalactic} \corr{sources are believed to give the dominant contribution to the high energy neutrino flux.} 
Assuming that the highest energy cosmic rays that we observe on Earth are typical of the entire cosmos and are of extragalactic origin, we estimate that they have a 
energy density of $\rho_{\mathrm{uhecr}}=3\times 10^{-19}$ erg/cm$^3$ 
above 1 EeV. 
Considering the typical evolution time of $T_H=10$ billion years, the corresponding energy losses of the universe are $W=\rho_{\mathrm{uhecr}}/T_H=9\times 10^{44}\,\mathrm{erg/(Mpc}^3\,\mathrm{yr)}$.
This cosmic ray population will be 
 in equilibrium if, in the reference volume of \corrca1 Gpc$^3$, there is a population of 
 900 $\gamma$-ray burst\corrca{s} and each one injects suddenly $10^{51}$ erg in cosmic rays; \corrca{an alternative would be that} there is a population of 150 active galactic nuclei, and each one radiates continuously $2 \times 10^{44}$ erg/s.
Interestingly, in both cases, the number of sources is reasonable and the presumed amount of energy emitted in cosmic rays corresponds to the visible electromagnetic output.

\paragraph*{Remark:--}
It is useful to examine the hypothesis that our Galaxy is a typical emitter of high energy neutrinos, even if the contrary could be true; to do it, we follow a very similar reasoning to that of Olbers' paradox. 
The density of galaxies \corrca{in the Universe} is about $n_{\mathrm{gx}}\sim 1$ Mpc$^{-3}$; 
\corrca{let us assume} that \corrca{each} of them, with typical size of $d\sim 10$ kpc, radiates cosmic rays and therefore neutrinos.
The intensity that we receive from the Milky Way is $\propto 1/(4\pi d^2)$. This is outshone by the integrated luminosity of those at a distance $D$ determined by 
$N_{\mathrm{gx}}/(4\pi D^2)=1/(4\pi d^2)$ where $ N_{\mathrm{gx}}=4\pi  D^3/3 \times n_{\mathrm{gx}}$: this condition  
corresponds to 2.5 Gpc, similar to the size of the entire Universe. 
The take-home message is the following one: the relative closeness of the Milky Way suggests not to dismiss lightly the possibility of a sizeable galactic emission of neutrinos - and other particles as well.

\paragraph*{\corr{Galactic sources:--}}
%\label{sec:CRgal}
For the galactic cosmic rays, it was noted that the gas around supernovae - named in astronomy ``supernova remnant'' - injects an amount of kinetic energy \corrca{larger by} one order of magnitude than the energy that the Milky Way loses in cosmic rays. 
It would thus be sufficient to convert \corrca{such amount of kinetic energy} in cosmic rays with an efficiency of $\sim 1/10$ to compensate \corrca{for the Milky Way} losses (Ginzburg-Syrovatskii hypothesis). 
Now, \corrca{in order to provide high energy neutrinos, an astronomical site has to be rich both in cosmic rays and in target material}: \corrca{consequently}, the association of supernova remnants and molecular clouds, the presumed sites of stellar formation, offers us the ideal condition for neutrino production. This applies to RX J1713.7-3946, a supernova remnant 1 kpc \corrca{away} from us, corresponding to a supernova seen in 393 AD by Chinese astronomers. 
Other hypotheses on galactic cosmic rays \corrca{sources} (or part of them) concern the central black hole, located in Sgr A*, and the surrounding regions,
that could extend \corrca{for} many kpc. %\corr{On the other hand, also the matter that is present in the Galactic disk and in the Galactic could provide the right target for cosmic rays. From their collisions high energy neutrinos can be created.}%Let us pass to the discussion of presumed extragalactic sources of cosmic rays. 

%{\Huge CONTROLLA i 4pi E METTI QUANTITATIVO - science ed io}

\subsection{Opportunities to see cosmic neutrinos} % pagg. 59-65
\label{sec:CRgammanu3}
Let us summarize the main facts on background events.  
Cosmic ray interactions in \corrca{the} atmosphere \corrca{produce} neutrinos at high energy. 
These \corrca{constitute} a diffuse, continuous and omnipresent flux, consisting mostly of muon (anti)neutrinos, with a spectrum distributed as $E^{-3.7}_\nu$ \corr{that gives a relevant contribution up to about} 100 TeV, with some uncertainty due to the prompt component \corrca- especially at high energies. 
Along with muons coming from above, these neutrinos are the irreducible background that we have to live with, when searching for cosmic neutrinos.

Keeping in mind these facts, one may search for, 
\begin{itemize}
\item \underline{sporadic sources.}
This was exploited to search for neutrinos associated to $\gamma$-ray burst in IceCube, and it has led to {\em exclude} that
a significant emission of cosmic neutrinos occurs  in the time around the 
$\gamma$-ray emission;
\item \underline{point sources,} possibly correlated with known objects; alternatively, one may search for multiple signals from the same spot in the sky. 
This case is especially interesting for galactic sources or for particularly brilliant extragalactic objects. 
Depending upon the angular resolution of the \rvw{neutrino telescope}, 
the energy threshold that can be used for this type of search can be even below TeV. However at present no such source is observed. 
\item \underline{diffuse sources.}
This would be the case, for instance, when many extragalactic sources contribute to an intense and measurable neutrino emission.  
In this case, the arrival direction of the neutrinos is almost isotropic, and one should rely mostly on the spectrum to search for a signal in \rvw{high-energy neutrino detectors}. 
One can cope with the atmospheric spectrum by using a relatively large energy threshold: typical choices in IceCube range from  
few 10 TeV to 200 TeV. IceCube observatory 
is collecting evidence for a cosmic neutrino signal falling in this class;
\item \underline{intermediate cases,} namely extended sources.
This could be the case concerning a galactic emission, due to the cosmic rays confined in the disk (or in its surroundings) \corrca{which then interact} with the gas contained in the Galaxy.\footnote{Note incidentally that a hypothetical emission from the halo would look, presumably, isotropic.} The current IceCube data do not exclude a minor ($\sim 10$\%) galactic component.
\end{itemize}
\rvw{The value of the angular resolution of the events $\delta\theta$ 
has an evident importance to really make astronomy.
In this regard, note  that the direction of the muons IceCube (in ice) 
can be identified with a precision of the order of 1 degree, while the other events 
are limited to 10 degrees at best. 
In water, these values are almost 10 times smaller, which implies a very important (quadratic!) 
improvement on the search window, i.e., 
on the solid angle $\pi\times \delta\theta^2$.}

It would be desirable to proceed with some firmer theoretical guidance; however, modeling neutrino emission at high energy is not easy and the expectations are unclear.  
We mention only few cases that stem from the previous considerations concerning the cosmic rays at the sources. 
Assuming that the high energy $\gamma$-rays observed from RX J1713.7-3946 are fully due to $\pi^0$ decays, one expects few muon neutrinos per km$^2$ per year above 1 TeV, that can be measured by km-class neutrino telescopes in the northern hemisphere.  
Similar considerations apply for the presumed neutrino flux from the Galactic center.
The extragalactic neutrino signal is potentially large but not precisely predicted.
One theoretical guide has been (and still it is) the so called Waxman-Bahcall bound, that follows  
by supposing that there are interactions in the source of ultra-high energy cosmic rays and assuming \corrca{energy equipartition in cosmic rays and neutrinos}. 
This condition fixes the normalization, and the distribution is postulated to obey a $E_\nu^{-2}$ law. 
A similar bound can be obtained considering the $\gamma$-rays that are emitted and then reprocessed; the measured intensity, once again, fixes the normalization of the expected neutrino flux.
This gives an isotropic flux, that can yield tens of events per km$^2$ per year above 30 TeV; the findings of IceCube, recalled just below, are a 
factor of few smaller.

\subsection{Observable events in neutrino telescopes} % pagg. 59-65
\label{sec:CRgammanu2}

%\centerline{\Huge\bf muons vs HESE}
At this point, it is necessary to clarify a bit more which types of signal can be detected in high energy neutrino telescopes. 
In fact, neutrinos and antineutrinos are not directly observable, but upon interaction with the matter they produce visible particles such as charged leptons, hadrons and photons. 

There are two types of signals of neutrinos that have been observed by IceCube, that  we  
introduce just below. Then we will mention two new types of signals that have not been seen yet.
%\par\medskip

%\noindent

\paragraph*{Induced muons:--}
The signal \corrca{produced by} \textit{muon neutrinos and/or antineutrinos} was considered already 60 years ago.
When they interact with the matter surrounding the neutrino telescope (ice, water or ground) they produce hadrons \corrca(that are readily stopped and absorbed before reaching the detector\corrca{) and (anti)}muons, that travel for a long distance instead and have \corrca{a} much \corrca{greater} chance of being seen.  
For a muon that propagates in water, with \corrca{initial} energy  
$E_\mu^{\mathrm{in}}$ and \corrca{final} energy $E_\mu^{\mathrm{fin}}$, 
an approximate formula for the path-length is:
\begin{equation}
\ell(E_\mu^{\mathrm{in}},E_\mu^{\mathrm{fin}})=2.5\;\mathrm{km}\times \log\left[\frac {1+E_\mu^{\mathrm{in}}/(\mbox{0.5 TeV})}{1+E_\mu^{\mathrm{fin}}/(\mathrm{0.5 TeV})} \right]
\end{equation}
where $E_\mu^{\mathrm{fin}}$ depends on the detector properties (e.g.\ threshold).
Thus the effective volume of the detector is a cylinder with basis the size of the neutrino telescope and with height the path-length of the muon; we may say that the matter around the detector acts as a converter for neutrinos. 
\corr{In other words, the effective volume is \corrca{larger} than the physical volume of the detector, in this case.}
The corresponding signal is called in several different manners: {\em induced muons, 
passing muons, throughgoing muons, upgoing muons, track signal,} etc. 
\rvw{The first neutrino telescopes \cite{spie}}
were located at more than 2 km \corrca{of} depth; below a certain zenith angle ($45^\circ$ and $60^\circ$ for CWI and KGF), the atmospheric muons \corrca{are} absorbed \corrca{in} the Earth and only \corrca{neutrino-}induced muons \corrca{produce} a signal.  
The most recent neutrino telescopes are not \corrca{as} deep, however they are able to tell the direction of the muon. 
In this manner, they can select the muons that arrive from below the horizon, \corrca{i.e.~that cannot be atmospheric in origin, as they are induced by neutrinos}. 
It should be noted that at \corrca{energies} above few 100 TeV the Earth is not transparent to neutrinos; therefore, the very high energy muons come preferentially from the regions just below the horizon, and \corrca{the closer their arrival direction is to the nadir, the smaller their flux gets}.

\paragraph*{HESE:--}
\corrca{The second type of signal is that of} \corr{High Energy Starting Events (HESE): this class \corrca{refers to} events with the interaction vertex inside the detector. 
For this reason the effective volume of the detector coincides with \corrca{its} physical volume. 
%while, in the previous case, the effective volume was greater than the physical volume. 
HESE are divided in two topologies of events: showers and tracks. 
Shower-like events \corrca{can be produced by} $\nu_\ele$ and $\nu_\tau$ \corrca{in the case of} charged current interactions, and \corrca{by} all neutrino flavors \corrca{in the case of} neutral current interactions. 
They are characterized by a poor angular resolution, about $10^\circ-15^\circ$ in ice and some degrees in water.\footnote{However, 
when used together with induced muons, they offer us a chance to test (and till now, 
they are consistent with) neutrino oscillations on cosmic scales, that will be discussed in the next section.}
The track-like events, instead, are only due to $\nu_\mu$, that interacts via charge\corrca{d} current interactions. 
They are characterized by a good angular resolution, of about $1^\circ$ in ice and sub-degree in water; the track-like events are therefore better for neutrino astronomy and in the search of connection between high energy neutrinos and known astronomical objects and sources of radiation (infrared, X, $\gamma$-rays,~...).}

\paragraph*{Double pulse:--}
A type of signal that is expected, but not observed yet, concerns \textit{tau \corrca{(anti)}neutrinos}. 
If a tau lepton of low energy decays in the detector, \corrca{its} corresponding signal is just as a\corrca{n} ordinary HESE signal; when the $\tau$ has \corrca{an energy of a fraction of PeV, however}, it is possible to distinguish the first energy release, consisting in \corrca{the hadrons produced at the interaction vertex of the $\tau$ (anti)neutrino}, from the second energy release, due to the decay of the \corrca{(anti)tau} {\em in a different place.} 
The signal is called {\em double pulse} and will be discussed further later on.\footnote{The principle is the same one adopted by detectors \rvw{based on photographic emulsions} - as was OPERA in Gran Sasso lab - but the scales are macroscopic - that suggested the witticism `cosmic OPERA' or 
`space OPERA' to indicate this signal.}

\paragraph*{Glashow resonance:--} 
Another expected but \corrca{unobserved signal} concerns \textit{electron antineutrinos.} 
These particles \corrca{can interact} with atomic electrons producing a real $\wbos$ boson, through the reaction $\nu_\ele+\ele\to\wbos$, when their energy is,
\begin{equation}E_\nu=m_\wbos^2/(2 m_\ele)=\mathrm{6.32\ PeV}\end{equation} 
\corrca{i.e.~more energetic than the neutrinos which have been observed so far}. 
This signal is called {\em Glashow resonance} and may allow us to probe the  
\corrca{most energetic} part of the cosmic neutrino spectrum. 

\paragraph*{Number of events:--}
The calculation of the expected number of neutrino signals $N_i$, given a flux of neutrinos of type $\ell$, is usually performed by means of {\em effective areas} $A_{\ell \to i}$. 
If the flux is continuous and the observation time is $T$, we can write symbolically:
 \begin{equation}N_i=\corr{4\pi} T \int dE_\nu \ A_{\ell \to i}(E_\nu)\  \frac{d\Phi_\ell}{dE_\nu}\end{equation}
Of course the effective area has \corrca{the} physical \corrca{units} of an area; it incorporates  the neutrino cross section, the number of target particles in the detector, the angular and the energy response, the cuts implemented, and it is a growing function of the energy of the incoming neutrino.
\corr{The flux measured by IceCube above 200 TeV, measured mostly by throughgoing muons, is still compatible, within 1$\sigma$, with an $E^{-2}$ spectrum, expected from a theoretical point of view:
\begin{equation}
E^{2} \frac{d \Phi_\mu}{dE_\nu} \simeq (0.7 \pm 0.3) \times 10^{-8} \frac{\mathrm{GeV}}{\rm cm^2 \ sec \ sr}
\end{equation} 
Assuming the pion decay as \corrca{the} main mechanism of \corrca{neutrino} production and considering that after neutrino oscillations $\Phi_\ele \simeq \Phi_\mu \simeq \Phi_\tau$ (as discussed in the next section) we can combine this flux and the IceCube effective areas, finding that the expected number of HESE events (all flavor) is about 7-8 events/year.}

\section{Neutrino transformations\label{sec:no}}
\subsection{Concept and basic formalism}

Neutrinos are produced in astronomical sites with some {\em initial composition of flavor}. For instance: 
\begin{enumerate}
\item in the Sun they are electron neutrinos $\nu_\ele$, since the matter is proton-rich and certain nuclear species transform according to $\prot\to \neut+\ele^++\nu_\ele$;
\item in supernovae, all types of neutrinos and antineutrinos are produced, due to all types of reactions, including the pair radiation $N+N\to N + N + \nu_\ell+\bar\nu_\ell$ with $\ell=\ele,\mu,\tau$; 
\item in cosmic ray collisions, around their sites of production, they are mostly $\nu_\ele,\nu_\mu,\bar\nu_\ele,\bar\nu_\mu$ (but not $\nu_\tau,\bar\nu_\tau$), since they are due to light meson decays such as $\pi^+\to \mu^++\nu_\mu$ followed by $\mu^+\to \ele^++\nu_\ele+\bar\nu_\mu$; 
\end{enumerate}

\agr{Once produced, however}, the neutrinos and antineutrinos with given flavor $\ell=\ele,\mu,\tau$ \agr{cannot propagate unperturbed, since they} {\em are not} states with \agr{definite} mass. 
They are rather superpositions of states with given mass, according to certain coefficients $U_{\ell j}$, known as leptonic mixing matrix; in formulae,
\begin{equation}
| \nu_\ell \rangle = \sum_{j=1}^3\ U_{\ell j}^*\ | \nu_j \rangle\quad\mathrm{and}\quad
| \bar\nu_\ell \rangle =  \sum_{j=1}^3\ U_{\ell j}\ | \bar\nu_j \rangle
\end{equation}  
The masses of the component neutrinos $|\nu_j \rangle$ are the same as the masses of the antineutrinos, and being non-zero but very small, below eV, they do not cause observable delays. The measurable effects are due to another circumstance:   
\agr{The propagating }neutrinos can be considered as de Broglie plane waves, that acquire \agr{different phases}
\begin{equation} |\nu_j, \, t, \,\vec{x}\rangle =      e^{- i \varphi_j}\,|\nu_j\rangle\ \quad\mathrm{with}\quad  \varphi_j =\frac{E_j\, t -  \vec{p}\cdot\vec{x}}{\hbar}\end{equation}
The energy of the individual states is
\begin{equation}
E_j= \sqrt{\vec{p}^2 + m_j^2}
\end{equation} 
where we set $c=1$ for convenience. 
The fact that the phases are different produces a non-zero overlap
\begin{equation}\langle  \nu_{\ell'}   |\nu_{\ell}, \, t,\, \vec{x}\rangle \quad\mathrm{with}\quad\ell\neq \ell'\end{equation}
namely, a transformation of the neutrinos (and also of the antineutrinos) in the course of their propagation. 
In certain cases, this effect has \corrca{an} oscillatory \corrca{behaviour} and this is the reason why (with an abuse of terminology) it is common to call these phenomena {\em neutrino oscillations}: 
see \cite{guido} for a thorough introduction.

% l'effetto della materia e l'equazione di propagazione

% gli effetti delle NC il problema non lineare non risolto per le SN

The most important quantities to describe the observable phenomena are the leptonic mixing matrix $U_{\ell j}$ and the phase differences of two neutrinos with different mass, that can be written as:
\begin{equation}
\begin{aligned}
\varphi_i-\varphi_j&=
(E_i-E_j)\,t=\frac{E_i^2-E_j^2}{E_i+E_j} t\\
&\approx \frac{m_i^2-m_j^2}{2 |\vec{p}|} t = \frac{m_i^2-m_j^2}{2 E} L
\end{aligned}
\end{equation}
where $L$ is the distance \corrca{covered during the} propagation, 
$L\approx c\, t = t$ and $E\approx |\vec{p}| c= |\vec{p}|$ (the first equality is due to the ultrarelativistic  condition, the second simply to the fact that we have set $c=1$). Below we will discuss the importance of these effects, chosing as an example the specific case of high energy neutrinos of cosmic origin.

The situation with solar neutrinos is slightly more complicated; in fact, the neutrinos are produced close to the center of the Sun, where there is an electronic \rvw{number density} of $\sim 100$ moles of electrons/cm$^3$. 
Electron neutrinos receive a further quantum phase (on top of those discussed just above) due to electron scattering: stated in other terms, the ordinary matter can be considered as a medium with a refraction index, that is different for electron neutrinos. 
This leads to interesting additional effects (called matter effect or MSW effect) that will be not discussed in details here. 

The case of supernova neutrinos is even more \corrca{complex}. 
In fact, in the production sites there are very large neutrino densities, and 
neutrinos can change their flavor simply crossing other neutrinos, as indicated by the following formula: 
\begin{equation}
\nu_\ele(\vec{p},\text{to obs.}) + \nu_\mu(\vec{q}%,\mathrm{ambient}
) \to \nu_\mu(\vec{p},\text{to obs.}) + \nu_\ele(\vec{q}%,\mathrm{ambient}
)
\end{equation}
(\agr{notice} that there is no exchange of momentum between the observed neutrino and the ambient one; only a change of flavor). 
For this reason, the problem of describing supernova neutrino transformation is non-linear, and to date, is considered still unsolved. 

\subsection{One application to neutrino astronomy}

%\paragraph{High energy neutrinos}
The simplest case, conceptually, is the one that concerns high energy neutrinos. 
In fact, it is plausible that they are produced in almost empty cosmic environments, so that we can use vacuum propagation formalism described above and can neglect the complications of the ambient matter. 
Moreover, the phases of oscillations are very large, e.g.,
\begin{equation}
\frac{\Delta m^2\ L}{2 E} =( 6 \times 10^6)  \frac{\Delta m^2}{7.37 \cdot 10^{-5} \mathrm{ eV}^2} \times \frac{L}{\mathrm{pc}} \times \frac{\mathrm{TeV}}{E}
\end{equation}
Thus, all oscillatory terms, when averaged over a certain energy range and/or distance of production, average to zero and can be omitted. 
This is called sometimes Gribov-Pontecorvo regime, or also {\em classical limit}.\footnote{In fact, the phases are inversely
 proportional to $1/\hbar$ and in the limit $\hbar\to 0$ the phases approach infinity.} 
 The probabilities become then simply a set of numbers,
 \begin{equation}
 P_{\nu_\ell \to \nu_{\ell'}}\equiv \lim_{t\to \infty} 
 | \langle \nu_{\ell'} |  \nu_{\ell} , t,\mathbf{x}\rangle |^2 =
 \sum_{i=1}^3 | U_{\ell i}^2 | | U_{\ell' i}^2 | 
 \end{equation}
 and in fact: 1) only three of these numbers are independent; 2) \corrca{none} of them is negligible.
There are three main implications of these formulae: 
\begin{enumerate}
\item All neutrino types are expected to arrive with similiar intensity on Earth; in fact, it is \corrca{customary} to assume the $1/3-1/3-1/3$ proportion in the investigation of cosmic high energy neutrinos, which is a particularly good approximation when neutrinos come from pion (and kaon) decay. 
\item Even if we do not have tau neutrinos at the production \corrca{site}, as in the most reasonable astrophysical mechanisms, we should have them after oscillations. 
The flux that has been observed by muon neutrinos is sufficient to derive the expectation of 0.2-0.3 event per year 
in the current dataset exceeding seven years, with conservative assumptions and within an uncertainty of about 20\%.
\item A similar conclusion applies \corrca{also} to electron antineutrinos, that are of interest for the production of real $\wbos$ bosons through the reaction with atomic electrons, $\bar\nu_\ele+ \ele\to \wbos$.  
This is true also in the case when the production of neutrinos happens via\\$\prot+\gamma(\mathrm{ambient})\to \Delta^+\to \neut +\pi^+$, even if the decay chain of the $\pi^+$ {\em does not } contain electron antineutrinos at the source. 
\end{enumerate}
Only the first circumstance, to date, has been verified by using the data of IceCube. 
The last conclusion could be evaded, if the cosmic neutrino spectrum has a cut \corrca{below} 6.32 PeV (i.e., when $\wbos$ bosons would be produced). 
However, the second prediction is particularly important to prove definitively the assumption that IceCube has seen neutrinos from cosmic sources; this will be subject of intensive studies in the coming months and years. 

At this point, it is useful to discuss more in details how it is possible observe tau neutrinos in IceCube. At relatively low energies, a tau neutrino decays so close to the production vertex that the event is practically indistinguishable from a shower event. Instead, at higher energies, the tau travels a bit before decaying, and this can make the event identifiable. Consider a tau neutrino that propagates $d=10$ meters
from the production vertex, where also energy is released; the \corrca{time separation of} the two pulses of energies
is of the order of $d/c=33$ ns, that is still within the capability of the detector.
If we require that this time is comparable with the lifetime of the tau, $t_\tau\times \gamma_\tau$ (where $t_\tau=0.29$ ps is the lifetime for a tau at rest and $\gamma_\tau$ the Lorentz factor) we have that $\gamma_\tau=E_\tau/m_\tau \approx 10^5$, thus, the energy of the tau is 200 
TeV. These considerations suggest that tau neutrinos above few hundreds TeV can lead to a {\em double pulse} signal in the individual phototubes. 
Note that the current observations of up-to-PeV muon neutrinos, along with three flavor neutrino oscillations, imply that these tau neutrinos exist for sure.

% HE ; il rapporto di flavor e la rivelazione di tau 

\section{Discussion\label{sec:di}}
Neutrino astronomy \corrca{deals with} a somewhat exotic particle. 
This implies that it has to \corrca{(heavily) rely on} other sciences, especially nuclear and particle physics; 
on the other hand, it allows us to investigate particle properties and nuclear transformations, especially when the relevant astrophysics is sufficiently well-known -- which is not always the case.

In these lecture notes, we have offered a brief introduction to this multidisciplinary branch of science, or, more precisely, we have  selected and discussed a few important topics that concern neutrino astronomy: solar neutrinos, supernova neutrinos, and high-energy neutrinos. 
For all these cases, we have argued that there is a significant amount of goals still to be achieved, and, therefore, of work that is still do be done.

\rvw{Many books and review works on neutrinos can help the Reader to deepen the related subjects, e.g., \cite{guido,r1,r2,r3} but note that those dedicated specifically to neutrino astronomy as~\cite{bahcall3,astro-book,spie}
are far less common to date.}

This discipline is one of the newest branches of astronomy, and it is still matter of important changes and surprises. 
As for other branches of astronomy at their inception, one would wish to proceed along a well-traced path, following standard steps in sequence as
follows:
\begin{enumerate}
\item Physics goal / hypothesis 
\item Precise predictions / expectations  
\item Principle of detection / telescope
\item First observation / discovery
\item Systematic study / measurements
\item Understanding 
\item New questions / anomalies 
\end{enumerate}
namely, from theory to observations and back to theory.

\corrca{However}, the history of astronomy, and even more of neutrino astronomy, \corrca{testifies} that Nature is not bound to conform {\em fully} to our wish of an ordered progress: the study of solar and atmospheric neutrinos has immediately revealed anomalies and has led us eventually to discover neutrinos oscillations.
Moreover, the high-energy neutrinos observed in IceCube surprised most (if not all) of us, and, after several years, the number of indubitable points in the discussion is rather limited.  

Thus, the only safe prediction is that neutrino astronomy will continue to give us a lot of fun, and we prepared these notes in the hope \corrca{of helping Readers to join} this enterprise or, at least, to share its joyful spirit.

{\small 
\subsection*{Acknowlegments}
%\newpage
We thank
M~Busso, 
A~Capone,
A~Esmaili, 
G~Fantini, 
W~Fulgione, 
A~Ianni, 
L~Latronico, 
P~Lipari, 
G~Pagliaroli, 
M~Spurio, 
V~Tretyak, 
D~Vescovi, 
FL~Villante, 
V~Zema
for discussions and  collaboration on these topics.
The work of AP has been partially supported with funds of  the European Research Council (ERC) under the European Uni\-on's Horizon 2020 research and innovation programme (Grant Number 646623).

}

%\appendix
%\addtocontents{toc}{\setcounter{tocdepth}{-1}}
\section*{Appendix: High energy  photons and neutrinos} % pagg. 66-70 + qualcosa altro
\begin{figure}
 \begin{center}
\includegraphics[width=.27\textwidth,keepaspectratio]{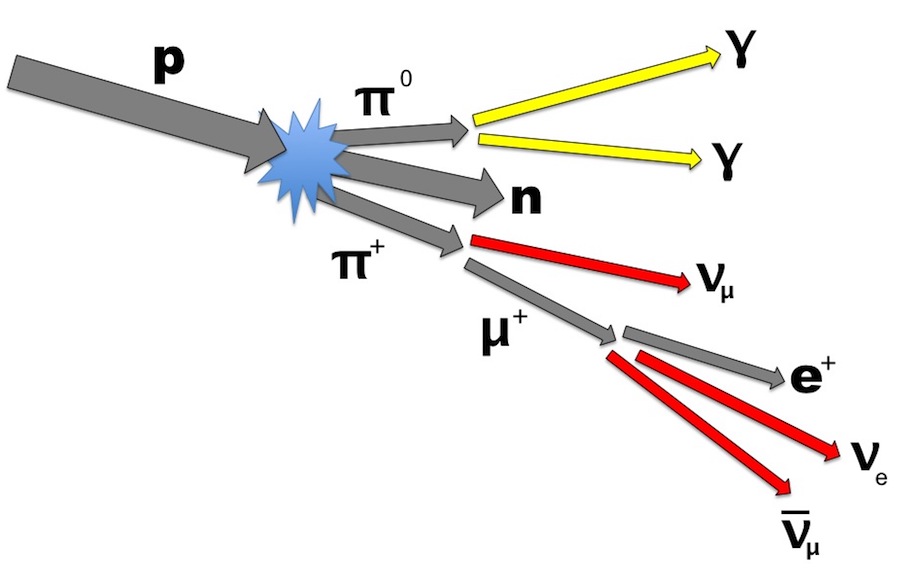}
\end{center}
\caption{Illustration of the connection between gamma-rays and neutrinos, that are produced when the primary cosmic rays collide with some target around their source.}
\label{illu} 
\end{figure}

\label{sec:CRgammanu1}
Photons are omnipresent in the universe and much better known than neutrinos in astronomy. It is not a surprise that there
are many possible connections with neutrino astronomy, either theoretical or observational. Here we discuss three possible different connections, that concern the case when these particle derive from cosmic rays collisions near to their source,
see figure~\ref{illu}.
%\begin{itemize}

%\item[a)]
\paragraph*{Production of secondary radiation:--}
\corr{There are two main mechanisms of high energy neutrino production. The first one is the collision between cosmic rays and target protons, that we denote as $\prot\prot$ interaction. The second one occurs when photons act as target for the cosmic rays to produce neutrinos and this is called the $p\gamma$ production mechanism. In both cases it is possible to establish connections between high energy neutrinos and photons.}
%as contrasted with the $\prot\prot$ production mechanism emphasized in the above considerations. 
The resonant production 
%\begin{equation}
$\prot+\gamma\to \Delta^+ $
%\end{equation}
is possible when $E_\gamma > (m_\Delta^2-m_\prot^2)/[2 (E_\prot  + p_\prot)]$; considering
e.g., protons of 1 PeV, this implies $E_\gamma>160 $ eV. Intense ultraviolet photon fields are in the vicinity of 
the supermassive black holes, that are harbored by the great majorities of the galaxies. Likewise, we have intense X-ray and $\gamma$-ray fields in several astronomical sites -- e.g., the jets 
of active galactic nuclei. Then, $\Delta^+$ can decay in $\pi^0 + \prot$ and $\pi^+ + \neut$, with a branching ratio 2:1 as dictated by isospin conservation.
This \corrca{does not hold in the case of} proton-proton collisions, since in this case an approximatively equal amount of $\pi^+,\pi^0$, and $\pi^-$ is produced. 
Thus, $\prot\gamma$ collisions lead to a larger amount of accompanying $\gamma$-rays. Another characteristic feature of this production mechanism is that the $\pi^+$ decay does not 
yield any electron antineutrino. A last aspect is that the shape of the spectrum of neutrinos depends upon the shape of the primary but also upon the photon distribution; for a $\prot\gamma$ production mechanism, deviations from a power law distribution and/or features in the spectrum are expected, \corr{whereas in the $\prot\prot$ mechanism the spectrum of neutrinos is expected to replicate the spectrum of primary protons}. \corrca{At the time of the writing}, it is unknown whether the main contributor to the cosmic neutrino population is due to $\prot\gamma$ or to $\prot\prot$.

%\item[b)] 
\paragraph*{Ordinary astronomy and neutrinos:--}
Then, there is one connection between neutrinos and photons that is in a sense obvious. %even if not less important. 
We have explored the Universe with photons and we may suppose - even before any theoretical elaboration - that some of the known astronomical objects are also sources of high energy neutrinos. Here, we examine the opportunities offered by the observations of photons of different energies: 
%\begin{itemize}
\newline \noindent $\bullet$
a \corrca{naive} hypothesis concerns the highest $\gamma$-ray sources (\corr{up to} some 10 TeV) that are currently known thanks to the Cherenkov telescopes (as HESS, Veritas, Magic, Milagro, etc).  As we discuss \corrca{hereafter}, this applies mostly to galactic objects;
\newline \noindent $\bullet$
when one considers the $\gamma$-ray sky as seen by \corrca{the} Fermi satellite below 100 GeV, the Milky Way is well \corrca{distinguishable,} but the main emission is from the rest of the sky, and it is dominated by the radiation emitted by the jets of the active galactic nuclei that point in our direction (blazars); 
\newline \noindent $\bullet$
ultraviolet and X-ray astronom\corrca{y} allow\corrca{s} us to explore the active galactic nuclei, and in particular their cores or the surrounding regions, 
but also supernova remnants or compact stellar objects in the Galaxy;
\newline \noindent $\bullet$
infrared astronomy allows us to see sites of intense stellar formation rich in dust, such as  star-forming and/ or star-burst galaxies; likewise, it is possible to see CO and H molecular emission from the molecular clouds of the Milky Way: these object 
surely have one key ingredient for efficient neutrino radiation, namely, the target for cosmic ray collisions; 
\newline \noindent $\bullet$ the exploration of the sites of most energetic processes of the universe is helped by 
radio astronomy as well, even if it concerns photons of comparably low energy.

%\item[c)] 
\paragraph*{Connecting directly high-energy $\gamma$ and neutrinos:--}
Let us consider the possibilities of a direct connection.
High energy neutrinos above TeV are necessarily produced along with $\gamma$-rays of similar energies, \corr{but the contrary is not true.}
We can use the observed $\gamma$-rays to predict a\corrca{n} upper bound for the neutrinos: in fact,  $\gamma$-rays are produced also by electromagnetic mechanisms, and, in this case, a part of 
them  is not associated with neutrinos. A direct connection requires two further conditions: on the $\gamma$-rays: 1)~they are not reprocessed in the source and 2)~they are not modified during the propagation.
The first condition depends upon the source; it is true for supernova remnants and typically for galactic sources (except possibly for micro-quasars and perhaps for the galactic center), while for extragalactic sources this condition should be examined case by case. 
The condition on the propagation instead rests on the efficiency of 
the pair conversion reaction 
$\gamma+\gamma\to \ele^++\ele^-$; the existence of the cosmic microwave background and of other photonic fields 
(especially the IR background, that is not precisely known) 
makes the universe opaque to $\gamma$-rays above 100 GeV and modifies also the galactic $\gamma$-rays with energies larger than a fraction of PeV. Thus, \emph{a direct connection between 
neutrinos above TeV and $\gamma$-rays is possible only for (some) galactic sources,
while for extragalactic sources the links are indirect and an extrapolation in energy is required if we want to compare the $\gamma$-ray and the neutrino spectra.}
Considering this type of galactic sources,  
it was shown \cite{saak} that we could receive a discernible signal in the neutrino telescopes\footnote{The calculation considered at least 1 muon per km$^2$ above 1 TeV; 
the muon signal, that will be discussed later on, is due 
to the conversion of neutrinos in the Earth -- througoing muon or induced muon signal.}
only if the $\gamma$-ray flux is larger than
\begin{equation} \tag{A.1}
%$$
I_\gamma(>\mathrm{10\ TeV})=(1-2)\times 10^{-13}\,\mathrm{cm}^{-2}\,\mathrm{s}^{-1}
%$$
\end{equation}
valid for few specific astrophysical sources. This   
draws a link with existing and future $\gamma$-ray telescopes.

\footnotesize
\tableofcontents
\end{document}